\documentclass[aps,prl,twocolumn,superscriptaddress,showpacs]{revtex4}
\usepackage{epsf}
\usepackage{amsmath,amssymb}
\usepackage{graphicx}
\usepackage{color}

\begin{document}

\title{Steering Bose-Einstein condensates despite time symmetry}

\date{\today}

\author{Dario Poletti}
\affiliation{Department of Physics and Centre for Computational
Science and Engineering,
National University of Singapore, Singapore 117542,
Republic of Singapore}
\author{Giuliano Benenti}
\affiliation{CNISM, CNR-INFM, and Center for Nonlinear and Complex Systems,
Universit\`a degli Studi dell'Insubria, Via Valleggio 11, 22100
Como, Italy}
\affiliation{Istituto Nazionale di Fisica Nucleare, Sezione di Milano,
Via Celoria 16, 20133 Milano, Italy}
\author{Giulio Casati}
\affiliation{CNISM, CNR-INFM, and Center for Nonlinear and Complex Systems,
Universit\`a degli Studi dell'Insubria, Via Valleggio 11, 22100
Como, Italy}
\affiliation{Istituto Nazionale di Fisica Nucleare, Sezione di Milano,
Via Celoria 16, 20133 Milano, Italy}
\author{Peter H{\" a}nggi}
\affiliation{Institut f{\" u}r  Physik, Universit{\" a}t Augsburg, Universit{\" a}tsstr. 1, D-86135 Augsburg, Germany}
\affiliation{Department of Physics and Centre for Computational
Science and Engineering,
National University of Singapore, Singapore 117542,
Republic of Singapore}
\author{Baowen Li}
\affiliation{Department of Physics and Centre for Computational
Science and Engineering,
National University of Singapore, Singapore 117542,
Republic of Singapore}
\affiliation{NUS Graduate School for Integrative Sciences and Engineering, 117597, Republic of Singapore}

\begin{abstract} 
A Bose-Einstein condensate in an oscillating spatially asymmetric potential is shown to exhibit a directed current for unbiased initial conditions despite time symmetry. This phenomenon occurs only if the interaction between atoms, treated in  mean-field approximation, exceeds a critical
value. Our findings can be described with a three-mode model (TMM). These TMM results  corroborate  well with a many-body study over a time scale which increases with increasing atom number. The duration of this time scale probes the validity of the used mean-field approximation. 
\end{abstract}

\pacs{05.60.-k, 03.75.Kk, 67.85.Hj}


\maketitle


The realization of Bose-Einstein condensates (BECs)~\cite{trento}
of dilute atomic gases has opened new possibilities for the investigation
of interesting physical phenomena induced by atom-atom interactions~\cite{morsch}.
For example, it was found that for a BEC in a double-well potential the tunneling can be
suppressed due to interaction-induced self-trapping~\cite{milburn,smerzi,albiez},
thus maintaining population inbalance between the two wells. A similar phenomenon
has been predicted also for a periodically driven BEC in a double
well~\cite{holthaus,weiss,luo,jiangbin,jiangbin2} and shows the
importance of interaction in the coherent control of quantum tunneling between wells.

In recent years, ample interest has arisen in the field of
directed transport in classical and quantum systems in absence of a
dc-bias, see~\cite{Hanggi, Hanggi2} and Refs. therein. In
particular it has been shown that by breaking space- and
time-inversion symmetries it is possible to achieve directed
(ratchet) transport~\cite{Flach,Flach2}. The study of the role of
interactions on the quantum ratchet transport, however, truly
remains {\it terra incognita}, with only a very few preliminary
studies available~\cite{Dario,Garcia,Flach3}. For instance,
in~\cite{Dario} it has been shown that atom-atom interactions in a
BEC can cause a finite current while the non-interacting system
would not exhibit directed transport; however, all these systems
possess a Hamiltonian that is both space- and time-{\it
asymmetric}.

With  this study, instead, we consider a 
non-dissipative, i.e. Hamiltonian system in which the
\emph{time-reversal} symmetry is not broken and the system is driven
smoothly. 
We show that, starting out from initial conditions symmetric both in momentum and
space, interactions, treated in the limits of a mean-field
approximation, induce directed current in a BEC. In a full many-body
analysis of the system, symmetry considerations imply that this
directed current is asymptotically decaying. Despite this fact we
show that a finite current persists over a time scale which
increases  with increasing atom number  in the condensate.


To start out, we consider a BEC composed of $N$ atoms confined in a toroidal trap of radius $R$
and cross section $\pi r^2$, subjected to the condition $r\ll R$, so that the motion is essentially one-dimensional.
The condensate is driven by a time-periodic potential, reading
\begin{equation}
V_{\rm Ext}(x,t)=\left[V_1\cos(x)+V_2\cos(2x+\phi)\right]\cos\left[\omega (t-t_0)\right],
\end{equation}
where $V_1$ and $V_2$ denote the potential depths of the two lattice 
harmonics, $\phi$ their relative phase,  and $\omega$ the angular frequency
of the oscillations of the perturbation. For $^6$Li and a trap of radius $R=2.5$ $\mu$m our used unit of time is $m R^2/\hbar=6.93\times 10^{-4}$ s. This potential can be readily reproduced in experiments~\cite{Weitz}.
The space symmetry ($x\to -x+\gamma$, with $\gamma$ constant~\cite{Flach}) is broken when $\phi\ne 0,\pi$.
The driving is symmetric under $t\rightarrow-t+2t_0$ inversion and in particular, for $t_0=0$, under $t\rightarrow -t$ time symmetry.

At zero temperature and as long as the number $N_0$ of condensate particles is much
larger than the number $\delta N$ of non-condensate ones, the evolution
of a BEC is well described by the
Gross-Pitaevskii (GP) equation~\cite{trento}.
Thus, for our system the evolution of the condensate
wave-function $\psi(x,t)$, normalized to $1$, is described by (taking $\hbar=m=1$) 
\begin{equation}
i\,\frac{\partial\psi(x,t)}{\partial t}=
\left[-\frac 1 2 \frac{\partial^2}{\partial x^2}+
g |\psi(x,t)|^2 + V_{\rm Ext}(x,t) \right]\psi(x,t),
\label{eq:evbec}
\end{equation}
where $x\in[0,2\pi)$ is the azimuthal angle of the torus, $g=8NaR/r^2$ the scaled strength of the nonlinear interaction (we consider the repulsive case, i.e., $g>0$) and $a$ is the $s$-wave scattering length for elastic atom-atom collisions.

As initial condition we use a wave-function which is symmetric both
in space, $x$, and in momentum, $p$, i.e. it is of the form
$\psi(x,0)=\sum_n a_ne^{i\alpha_n}\cos(nx)$, where $n\in
\mathbb{N}$ and $a_n,\alpha_n\in\mathbb{R}$. For simplicity, and
without loss of generality we choose the initial condition to be:

\begin{equation}
\psi(x,0)=\frac 1 {\sqrt{2}}
\left[\frac 1 {\sqrt{2\pi}}+\frac{e^{i\alpha}}{\sqrt{\pi}}\cos(x)\right],
\label{eq:initialconditions}
\end{equation}
where $\alpha$ is the relative phase between the homogeneous part and
the last term which corresponds to a cosinusoidal modulation in the initial
density of particles.
While in the non-interacting ($g=0$) case the asymptotic current is
zero for any initial condition that is symmetric in $x$ and $p$ , we
show that this is not the case if $g\ne 0$.

\begin{figure}[!h]
\includegraphics[width=8.0cm]{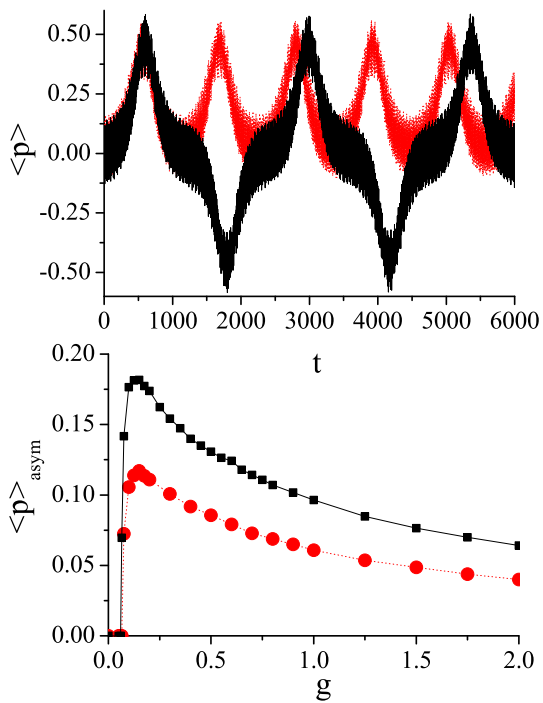}
\caption{(color online) Upper panel: Momentum expectation $\langle p \rangle$
{\it vs.} time  $t$ for $\alpha=0$ (dark black curve) and
$\alpha=\pi/2$ (light red curve). Used parameter values are:
$g=0.2$, $V_1=V_2=2$, $\phi=\pi/2$, 
$\omega=10$, $t_0=0$. Lower panel: Asymptotic
time-average momentum ${\langle p \rangle}_{\rm asym}$ versus the
interaction strength $g$ for $\alpha=\pi/2$. Data are obtained from
the GP equation (\ref{eq:evbec}) (black squares) or from the
TMM-Ansatz (\ref{eq:TMMansatz}) (red circles). In the first 
case $g^\star\approx 0.065$ and $g_{\rm opt}\approx 0.15$,
in the latter $g^\star\approx 0.070$ and $g_{\rm opt}\approx 0.15$.}
\label{fig:figK2}
\end{figure}

As depicted with Fig.~\ref{fig:figK2} (top panel), our model may exhibit a
non-zero asymptotic current, $\langle p\rangle_{\rm asym}=\lim_{t\to
\infty}\frac{1}{t} \int_0^t \langle p \rangle(t') dt'$, when $g\ne
0$ and $\alpha\ne 0$ with $\langle p\rangle (t')=-i\int_0^{2\pi}\psi^*(x,t')\frac{\partial} {\partial x}\psi(x,t')dx$. The asymptotic time-averaged
momentum ${\langle p \rangle}_{\rm asym}$ {\it vs.} $g$, Fig.~\ref{fig:figK2}\/ (bottom panel), clearly shows that (i) there is a
finite threshold value $g^\star$ such that ${\langle p \rangle}_{\rm
asym}\ne 0$ when $g>g^\star$ and (ii) there exists an optimal value
$g_{\rm opt}$ that maximizes the current.

\begin{figure}[!h]
\includegraphics[width=8.0cm]{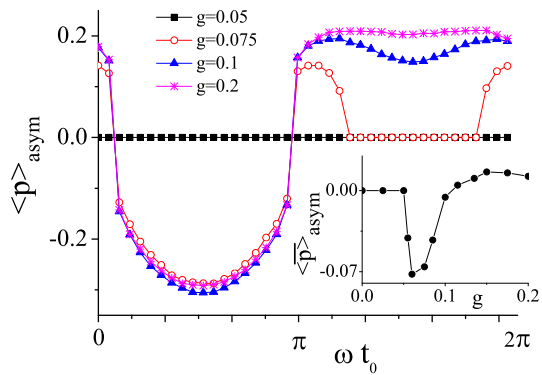}
\caption{(Color online) Asymptotic time-averaged momentum $\langle p
\rangle_{\rm asym}$ versus $\omega t_0$ for $g=0.05$ (filled black
squares), $g=0.075$ (empty red circles), $g=0.1$ (filled blue
triangles), and $g=0.2$ (pink asterisks). Inset: asymptotic current
averaged over $t_0$, ${\bar{\langle p \rangle}}_{\rm asym}$, as a
function of the interaction strength $g$. Parameter values: $V_1=V_2=2$,
$\phi=\pi/2$, $\omega=10$, $\alpha=\pi/2$.} 
\label{fig:t0}
\end{figure}

The asymptotic time-averaged current
is shown as a function of the initial phase $\omega t_0$ of the driving field
in Fig.~\ref{fig:t0}.
We can deduce that there is a strong asymmetry around
$\omega t_0=\pi$, in particular for the coupling strength  $g=0.075$.
As a result, the directed current survives even after averaging over
$t_0$ (cf. the inset of Fig.~\ref{fig:t0}).

\begin{figure}[!h]
\includegraphics[width=8.0cm]{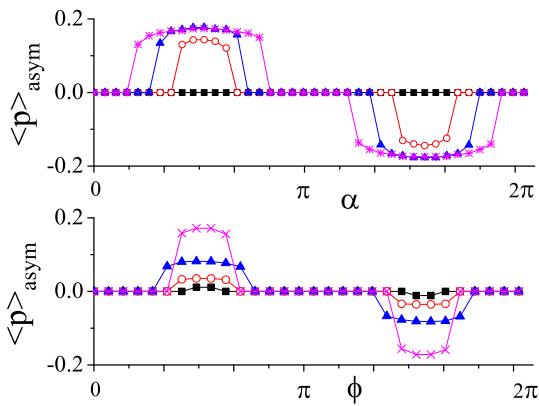}
\caption{(Color online) Top panel: Asymptotic time-averaged momentum $\langle p \rangle_{asym}$ versus $\alpha$ for $g=0.05$ (full black squares), $g=0.075$ (empty red circles). $g=0.15$ (full blue triangles) and $g=0.2$ (pink crosses). 
The other parameters values 
are $V_1=V_2=2$, $\phi=\pi/2$, $\omega=10$ and $t_0=0$.
Bottom panel: Asymptotic time-averaged momentum $\langle p \rangle_{\rm asym}$ versus $\phi$ for $V_1=V_2=0.2$ (full black squares), $V_1=V_2=0.5$ (empty red circles). $V_1=V_2=1$ (full blue triangles) and $V_1=V_2=2$ (pink crosses). Other parameters values are $g=0.2$, $\alpha=\pi/2$, $\omega=10$ and $t_0=0$.}
\label{fig:phase}
\end{figure}

The asymptotic time-averaged momentum is depicted as a function of
the relative phase $\alpha$ in Fig.~\ref{fig:phase} (top panel). We  note that
for certain intervals of the values of $\alpha$ it is possible to
obtain a non-zero current, and these regions widen as $g$ increases.
There is a symmetry for $\alpha\to 2\pi-\alpha$ and hence the
average over all $\alpha$ yields zero current.

\begin{figure}[!h]
\includegraphics[width=8.0cm]{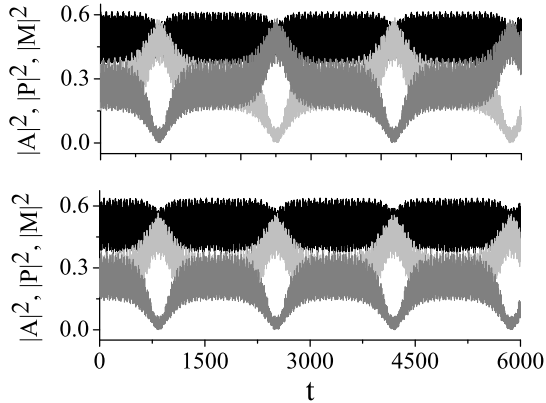}
\caption{Population of the three levels, $|A|^2$ (black), $|P|^2$(light gray), and $|M|^2$
(gray) which carry momentum 0,+1, and -1, respectively, for
$\alpha=0$ (top) and  $\pi/2$ (bottom), at $g=0.2$, 
$V_1=V_2=2$, $\phi=\pi/2$, $\omega=10$,
$t_0=0$.}
\label{fig:3mod}
\end{figure}


This $\alpha\to (2\pi-\alpha)$ symmetry can be reasoned as follows:
With $t_0=0$, let
$t\to \tau=-t$ and perform complex conjugation in the GP equation
(\ref{eq:evbec}), to obtain
\begin{equation}
i\,\frac{\partial\tilde{\psi}(x,\tau)}{\partial \tau}=
\left[-\frac 1 2 \frac{\partial^2}{\partial x^2}+
g |\tilde{\psi}(x,\tau)|^2 + V_{\rm Ext}(x,\tau) \right]\tilde{\psi}(x,\tau),
\end{equation}
where $\tilde{\psi}(x,\tau)\equiv \psi^\star(x,-\tau)$.
Therefore, if $\psi(x,t)$ is a solution of the GP equation, so is
$\tilde{\psi}(x,\tau)$. It follows that
$\langle \tilde{p}(\tau) \rangle$ of the wave function $\tilde{\psi}(x,\tau)$
is such that $\langle \tilde{p}(\tau) \rangle=-\langle p(t)\rangle$,
where $\langle p(t) \rangle$ is the momentum expectation of $\psi(x,t)$.
For the initial condition (\ref{eq:initialconditions}),
$\psi(x,0)\to \psi^\star(x,0)=\tilde{\psi}(x,0)$ when
$\alpha\to (2\pi-\alpha)$. We thus conclude that
$\langle p \rangle_{\rm asym}(\alpha)=-\langle p \rangle_{\rm asym}
(2\pi-\alpha)$.
Clearly, if $\psi(0)\in \mathbb{R}$
[$\alpha=0,\pi$ in Eq.~(\ref{eq:initialconditions})]
then $\langle p \rangle_{\rm asym} =0$.
While in the non-interacting case the current is
asymptotically vanishing for any initial condition, complex or
real~\cite{Flach2}, in the $g\ne 0$ case
a directed current can arise when
$\alpha\ne 0,\pi$.


In Fig.~\ref{fig:phase} (bottom panel) we show that a directed current 
can be generated only when the potential is spatially asymmetric,
that is, $\phi\ne 0,\pi$. Note that non-zero current is obtained
and remains stable in two $\phi$-regions around the maximum symmetry
violation values $\phi=\pi/2,3\pi/2$.


From our numerical studies we notice that the only states that are
significantly excited during the dynamical evolution of the system
assume the momenta $0$, $+1$ or $-1$. We
hence attempt to reproduce our present results, at least on a
qualitative level,  with the use of a mean-field three-mode model (TMM). We use the Ansatz
\begin{equation}
\psi(x,t)=\frac{1}{\sqrt{2\pi}}\left[A(t)+P(t)e^{i\;x}+
M(t)e^{-i\;x}\right],
\label{eq:TMMansatz}
\end{equation}
where $A(t)$, $P(t)$ and $M(t)$ are time-dependent, complex
coefficient such that $|A|^2+|P|^2+|M|^2=1$.
Using (\ref{eq:TMMansatz}) we find the effective mean-field
Hamiltonian of the TMM:
\begin{equation}
\begin{array}{c}
H_{\rm eff}=\int_0^{2\pi} dx\;
\psi^\star(x,t)
\left[-\frac 1 2 \frac{\partial^2}{\partial x^2}+
\frac{g}{2} |\psi(x,t)|^2 \right.
\\
\left. + V_{\rm Ext}(x,t) \right]
\psi(x,t),
\end{array}
\end{equation}
together with the equations of motion
\begin{equation}
\left\{\begin{array}c
i\dot{A}=\frac{g}{2\pi}\left( (|A|^2+2|P|^2+2|M|^2)A+2A^*PM  \right)\\
 +\frac{K}{2}(P+M)\cos(\omega t),\\
\\
i\dot{P}=\frac{P}{2}+\frac{g}{2\pi}\left( (|P|^2+2|A|^2+2|M|^2)P+A^2M^* \right)\\
 +\frac{K}{2}(A+iM)\cos(\omega t),\\
\\
i\dot{M}=\frac{M}{2}+\frac{g}{2\pi}\left((|M|^2+2|A|^2+2|P|^2)M+A^2P^* \right)\\
 +\frac{K}{2}(A-iP)\cos(\omega t).
\end{array}\right. \label{EOM3M}
\end{equation}

This TMM qualitatively reproduces the behavior of the asymptotic
time-averaged current, see Fig.~\ref{fig:figK2} (bottom panel). In
Fig.~\ref{fig:3mod} we depict the evolution of the
populations $|A|^2$, $|P|^2$, and $|M|^2$ of the three modes. For
$\alpha=0$ ($\alpha=\pi/2$) the time-averaged current 
is zero (non-zero).


The obtained results may appear surprising if one observes that
the full second-quantized quantum many-body operator is linear, obeying
time-reversal symmetry, and symmetry arguments necessarily imply a vanishing asymptotic current~\cite{Flach2}.
Actually there occurs no contradiction because the two limits $t\to
\infty$ and $N\to \infty$ do not commute: a directed current is
obtained {\it only} in the mean-field limit in which we let first
$N\to\infty$ and then $t\to\infty$. The issue for a physical system
in which both the number of particles and the time of the
experimental run are finite constitutes the vindication of the regime of
validity  of the used nonlinear GP equation. For this purpose, we
next study the second quantized  many-body problem within the three-mode
approximation, being also accessible to numerical investigation.

The second-quantized quantum many-body Hamiltonian reads:
\begin{equation}
\hat{H}_{\rm eff}=\int_0^{2\pi} dx\;
\hat{\psi}^\dagger
\left[-\frac 1 2 \frac{\partial^2}{\partial x^2}
+
\frac{g}{2N}
\hat{\psi}^\dagger
\hat{\psi}
 +  V_{\rm Ext}(x,t)\right]\hat{\psi},
\label{3bham}
\end{equation}
where, within the three-mode approximation, 
\begin{equation}
\hat{\psi}^\dagger=\frac{1}{\sqrt{2\pi}}(\hat{a}^\dagger+
\hat{p}^\dagger e^{ix} + \hat{m}^\dagger e^{-ix}),
\end{equation}
$\hat{a}^\dagger$, $\hat{p}^\dagger$, and $\hat{m}^\dagger$
denote the Boson creation operators for the states with momenta
$0$, $+1$, and $-1$. These operators satisfy the commutation relations
$[\hat{a},\hat{a}^\dagger]=1$,
$[\hat{p},\hat{p}^\dagger]=1$,
$[\hat{m},\hat{m}^\dagger]=1$ and operators corresponding to different modes
commute. The particle conservation reads
$\langle \hat{a}^\dagger\hat{a}+\hat{p}^\dagger\hat{p}+
\hat{m}^\dagger\hat{m} \rangle=N$, where $N$ is the total number of particles.
As initial condition, we consider $N$ atoms in the state
\begin{equation}
\frac{1}{\sqrt{N!}}\left[\frac{\hat{a}^\dagger}{\sqrt{2}} + \frac{e^{i\alpha}\left(\hat{p}^\dagger + \hat{m}^\dagger\right)}{2}\right]^N\,|0\rangle.
\label{3modesquantized}
\end{equation}

\begin{figure}[!h]
\includegraphics[width=8.0cm]{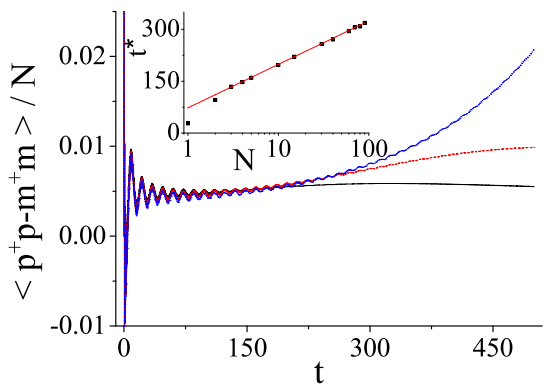}
\caption{(Color online) Population inbalance divided by $N$ between
levels with $p=+1$ and $p=-1$ {\it vs.} time $t$. From bottom to
top: $N=10$, $80$, and mean-field TMM, for $g=0.2$, 
$V_1=V_2=2$, $\phi=\pi/2$,
$\omega=10$, $t_0=0$, $\alpha=\pi/2$. Inset: $t^*$ versus $N$
(squares) and numerical fit $t^\star= A+B \ln N$, with $A\approx 73$
and $B\approx 54$.} \label{fig:NVMB}
\end{figure}

We studied numerically  the evolution of the $N$-body system
governed by the second-quantized three-mode Hamiltonian (\ref{3bham}).
The normalized population inbalance
between levels with momenta $+1$ and $-1$ is depicted in
Fig.~\ref{fig:NVMB}. The TMM holds up to
increasingly longer times $t^\star$ as $N$ increases.
We estimate $t^\star(N)$ as the time instant at which the $N$-body population
inbalance deviates by more than $1\%$ from the mean-field treatment value:
Our numerical data are well fitted by a logarithmic dependence,
$t^\star \propto \ln N$, cf. inset of Fig.~\ref{fig:NVMB}.

To check the stability of the condensate, we have 
diagonalized the single-particle reduced density matrix, whose
matrix elements read $\langle  \hat{x}^\dagger \hat{y} \rangle$, with 
$\hat{x},\hat{y}$ being $\hat{a},\hat{p}$ or $\hat{m}$. According to \cite{Penrose}, we have a 
BEC if this matrix possesses one eigenvalue of order $N$. 
Defining $\tilde{t}$ as the time for which the condensed state is highly populated, we have found numerically that $\tilde{t}\propto N^{0.45}$. This means that for large $N$ and times $t$ obeying $\tilde{t} >t> t^*$ the mean-field approach no longer provides an accurate description of the many-body system despite the presence of a condensate. 

In conclusion, we have shown that in BEC the interaction plays a
prominent role in inducing long lasting currents, when
otherwise impossible. We note that BECs in
toroidal traps have  recently been realized in different
experiments~\cite{Stamper,Olson,Phillips,Foot}. The initial
condition (\ref{eq:initialconditions}) can be implemented by
properly exciting the ground-state of a condensate in a toroidal
trap, and the phase $\alpha$ can be tuned by letting the excited
condensate  evolve over a given time span. The interaction strength
can be changed by tuning an external magnetic field~\cite{Donley}. Extrapolating the dependence given in Fig.~\ref{fig:NVMB} of $t^\star$
on the number $N$ of condensate particles, we obtain $t^\star\approx
800$ (in an in situ experiment this corresponds a time scale $0.1$
s) for $N\approx 5\times 10^5$ particles~\cite{Phillips}. Changing the number of condensed particles~\cite{Raizen} opens the
possibility to explore {\it in situ} the range of validity of the
GP description via the measurement of our  predicted,
intriguing directed transport.

We acknowledge fruitful discussions with A. Mouritzen and A. Smerzi. This work is supported in part by the Academic Research Fund, WBS grant no. 150-000-002-112 (B.L.).
Support by the German Excellence Initiative via the ``Nanosystems
Initiative Munich (NIM)'' (P.H.) is gratefully acknowledged as well.

\end{document}